# The Vertebrate Breed Ontology: Towards effective breed data standardization

Kathleen R. Mullen (1), Imke Tammen (2), Nicolas A. Matentzoglu (3), Marius Mather (4), James P. Balhoff (5), Elizabeth Esdaile (6), Gregoire Leroy (7), Carissa A. Park (8), Halie M. Rando (9), Nadia T. Saklou (10), Tracy L. Webb (10), Nicole A. Vasilevsky (11), Christopher J. Mungall (12), Melissa A. Haendel (1), Frank W. Nicholas (2), Sabrina Toro (1)

(1) Department of Genetics, University of North Carolina at Chapel Hill, Chapel Hill, NC, USA.
(2) Sydney School of Veterinary Science, The University of Sydney, Sydney, NSW, Australia
(3) Semanticly Ltd, Athens, Greece.
(4) Sydney Informatics Hub, The University of Sydney, Sydney, NSW, Australia.
(5) Renaissance Computing Institute, University of North Carolina at Chapel Hill, Chapel Hill, NC, USA.
(6) Veterinary Genetics Laboratory, School of Veterinary Medicine, University of California, Davis, CA, USA. Current affiliation is Department of Clinical Sciences, Colorado State University, Fort Collins, CO, USA.
(7) Animal Production and Health Division, Food and Agriculture Organization of the United Nations, Rome, Italy.
(8) Department of Animal Science, Iowa State University, Ames, IA, USA.
(9) Department of Computer Science, Smith College, Northampton, MA, USA.
(10) Department of Clinical Sciences, Colorado State University, Fort Collins, CO, USA.
(11) Data Collaboration Center, Critical Path Institute, Tucson, AZ, USA.
(12) Lawrence Berkeley National Laboratory, Berkeley, CA, USA

List of abbreviations: ADPKD, autosomal dominant polycystic kidney disease; COB, Core Ontology for Biology and Biomedicine; DAD-IS, Domestic Animal Diversity Information System; EHRs, electronic health records; EMBL-EBI, European Molecular Biology Laboratory - European Bioinformatics Institute; FAO, Food and Agricultural Organization; FIFe, Fédération Internationale Féline; GCCF, Governing Council of the Cat Fancy; ID, identifier; KA, knowledge acquisition; LBO, Livestock Breed Ontology; MIRO, Minimum Information for the Reporting of an Ontology; NCBI, National Center for Biotechnology Information; NCBITaxon, NCBI Taxonomy; OBO, Open Biological and Biomedical Ontology; ODK, Ontology Development Kit; OLS, EMBL-EBI Ontology Lookup Service; OMIA, Online Mendelian Inheritance in Animals; OMO, OBO Metadata Ontology; ORCID, Open Researcher and Contributor ID; OWL, Web Ontology Language; PKD1, polycystin 1, transient receptor potential channel interacting; QA, Quality Assurance; QC, Quality Control; REFR, Rare and Exotic Feline Registry; RO, Relation Ontology; SRD, scope, requirements, development community; TICA, The International Cat Association; VBO, Vertebrate Breed Ontology; VeNom, Veterinary Nomenclature; VETSCT, The Veterinary Extension of Systematized Nomenclature of Medicine – Clinical Terms; WCF, World Cat Federation.

Corresponding author: Kathleen R. Mullen, Department of Genetics, 120 Mason Farm Rd., Chapel Hill, NC, 27599, katie@tislab.org.

Acknowledgements: The authors wish to thank Leslie Lyons for sharing her list of cat breeds and acknowledge the contributions and support of Amos Bairoch, Rebecca Bellone, David Brodbelt, Zhiliang Hu, and the team at the Monarch Initiative. The authors thank Cellosaurus for the early adoption of VBO. This work was supported by NIH Office of the Director Grant #5R24OD011883 for the Monarch Initiative, the Ronald Bruce Anstee bequest to the Sydney School of Veterinary Science for the Anstee Hub for Inherited Diseases in Animals and the University of North Carolina at Chapel Hill Department of Genetics. K. Mullen was supported by the National Institute Of Arthritis And Musculoskeletal And Skin Diseases of the National Institutes of Health under Award Number K12AR084226. N. Saklou was supported by the National Institute of Health National Center for Advancing Translational Sciences (NCATS) U01TR002953-05. T. Webb was partially supported by NIH/NCATS Colorado CTSA Grant Number UM1 TR004399. C. Park was supported by the United States Department of Agriculture National Institute of Food and Agriculture (grant number 2022-67015-36217). Contents are the authors' sole responsibility and do not necessarily represent official NIH views.

Conflict of interest: M. Haendel is the founder of Alamya Health, which has no relevance to the work presented here.

# Abstract

<u>Background</u> - Limited universally-adopted data standards in veterinary medicine hinder data interoperability and therefore integration and comparison; this ultimately impedes the application of existing information-based tools to support advancement in diagnostics, treatments, and precision medicine.

<u>Hypothesis/Objectives</u> - A single, coherent, logic-based standard for documenting breed names in health, production, and research-related records will improve data use capabilities in veterinary and comparative medicine.

<u>Animals</u> - No live animals were used.

<u>Methods</u> - The Vertebrate Breed Ontology (VBO) was created from breed names and related information compiled from the Food and Agriculture Organization of the United Nations, breed registries, communities, and experts, using manual and computational approaches. Each breed is represented by a VBO term that includes breed information and provenance as metadata. VBO terms are classified using description logic to allow computational applications and Artificial Intelligence–readiness.

<u>Results</u> - VBO is an open, community-driven ontology representing over 19,500 livestock and companion animal breed concepts covering 49 species. Breeds are classified based on community and expert conventions (e.g., cattle breed) and supported by relations to the breed's genus and species indicated by National Center for Biotechnology Information (NCBI) Taxonomy terms. Relationships between VBO terms (e.g., relating breeds to their foundation stock) provide additional context to support advanced data analytics. VBO term metadata includes synonyms, breed identifiers/codes, and attributed cross-references to other databases.

<u>Conclusion and clinical importance</u> - The adoption of VBO as a source of standard breed names in databases and veterinary electronic health records can enhance veterinary data interoperability and computability.

## Introduction

Precision medicine holds great promise for advancing disease characterization and targeted drug therapies.[1–7] The success of individualized patient care relies on the availability of data, including molecular mechanisms of disease, genomic profiling, and pharmacogenomics,[3,6] from research databases and electronic health records (EHRs). However, veterinary data are rarely represented using universally accepted standards, making computability and integration challenging. Standards ensure consistency across sources and foster data interoperability, integration, and comparison, therefore supporting precision medicine applications. For example, representing a "gene" using a National Center for Biotechnology Information (NCBI) gene identifier (ID) makes a data entry unambiguous, and related information referring to the same ID can then be confidently integrated and compared. Standardized terminologies have been widely adopted in model organism and human databases and EHRs.[8] Ontologies are systematic representations of knowledge, which have been accepted as gold standards for their clearly defined concepts and rich metadata, including synonyms. Ontologies leverage description logic (i.e., computable relationships between terms within and across ontologies and terminologies), providing increased context and interoperability with a broad range of data during analytics.[9]

The concept of "breed" is essential for veterinary genomics; it provides a framework for understanding inherited traits, disease susceptibility, and genetic diversity within and across populations. Breed names are often embedded in free text notes or flat lists from practice management software. These lists vary between data sources and are rarely connected (e.g., there is rarely any indication that "German Shepherd", "Alsatian", and "Deutscher Schäferhund" available in different lists refer to the same dog breed). Some breed name standards exist (e.g., The Veterinary Extension of Systematized Nomenclature of Medicine – Clinical Terms (VETSCT)[10], The Livestock Breed Ontology (LBO)[11], and the breed-name component of Veterinary Nomenclature (VeNom) Codes).[12] However, these standards are either limited in scope (e.g., LBO is limited to livestock breeds), impose restricted licenses, or lack an ontological foundation; this makes these existing standards impractical for computational analyses and for a wide range of uses including animal husbandry and veterinary clinical research. An open-source standard that reconciles all breed names and information and their provenance is needed to ensure global data interoperability, which will in turn allow for advances in precision medicine and learning healthcare and inform care and breeding best practices.

Here we introduce the Vertebrate Breed Ontology (VBO) as a comprehensive source for breed names and metadata across all vertebrate animals, including livestock and companion animals. Similar to other Monarch Initiative[13] ontologies, VBO is open, community-driven, and provides linkouts to resources and supporting information. We describe VBO's creation, and how its use provides a powerful tool for breed-related data interoperability and can support data integration and precision medicine.

## Materials and Methods

**Scope of VBO**

Since the concept of "breed" is not clearly defined among communities, we took a broad approach when defining "vertebrate breed". We created VBO terms representing breed concepts as reported by any international breed organization and/or other expert communities.

**Sources for breeds and related information**

Through active collaborative engagements with international organizations (e.g. the Food and Agricultural Organization (FAO)), communities, and experts, we gathered lists of breeds and related breed information from relevant sources. A full list of these sources can be found in the VBO documentation.[14] Each breed source has specific goals; for instance, FAO's Domestic Animal Diversity Information System (DAD-IS)[15] aims to monitor breed diversity and sustainable use around the world, while canine, feline, and livestock registration bodies focus on breed standards and documentation.[16–18] Information included in breed lists is specific to the sources and can be contradictory. We incorporated all information without discrimination and included provenance and robust attribution to these sources.

**VBO content**

We manually reviewed and curated the breed lists to (1) group information related to the same breed under the same VBO term; (2) create VBO term names based on the most commonly used breed name and common species name, ensuring term label uniqueness (see VBO documentation[14]); and (3) map breeds to their corresponding NCBI Taxonomy (NCBITaxon) record (representing the species). VBO terms were integrated within the NCBITaxon hierarchy using the *is_a* relation. Relations between breeds (e.g., indicating the breed's foundation stock) were also manually curated based on breed information from the sources.

To facilitate ontology browsing and use, we created high-level grouping terms, such as 'dog breed', 'cattle breed', etc., that were logically defined based on their NCBITaxon parentage (subspecies, species, genus, or family). VBO terms were automatically classified under these high-level terms using a description logic reasoner. Rich metadata and cross-references to other terminologies and databases, including their provenance, were recorded for each VBO term.

**Creation of VBO using the Ontology Development Kit (ODK)**

The ODK[19] provides a framework for creating ontologies, including both executable workflows for managing ontologies, such as release workflow and continuous integration, as well as ontology-processing tools such as ROBOT.[20] The ODK is used to automatically check VBO for errors whenever changes are proposed (e.g., new classes are added) and to release new ontology versions. The ontology is managed and openly available on GitHub at https://github.com/monarch-initiative/vertebrate-breed-ontology. VBO has been accepted into the Open Biological and Biomedical Ontology (OBO) Foundry, a community of ontology developers committed to building open-source ontologies under a set of guiding principles (https://obofoundry.org/).[21]

**VBO maintenance**

Most external breed sources do not have unique and permanent identifiers that allow for a robust automated workflow for incorporating and synchronizing updated information. Therefore, VBO is currently mostly maintained based on user reviews and requests for changes or additions of new breeds submitted to the VBO GitHub repository (https://github.com/monarch-initiative/vertebrate-breed-ontology/issues). More information about ontology content and maintenance can be found in Table 1, including minimum information for the reporting of an ontology (MIRO).[22]

## Results

**VBO is a standard for breeds and breed information**

VBO represents the concept of breed broadly in order to cover information from multiple sources. In accordance with FAO,[23] we defined vertebrate breed as a taxonomic entity representing a population of vertebrate animals that share specific characteristics (such as traits, behavior, genetics) and/or for which cultural or geographical separation has led to the general acceptance of its separate identity. Breeds are formed through genetic isolation and either natural adaptation to the environment or selective breeding, or a combination of the two. Family, genus, and species are represented by NCBITaxon entities. Therefore, VBO is integrated within the NCBITaxon hierarchy (Figures 1 and 2), and VBO terms are ontological subclasses of these NCBITaxon entities.

VBO is openly available and browsable in ontology browsers such as the EMBL-EBI Ontology Lookup Service (OLS, https://www.ebi.ac.uk/ols4/ontologies/vbo), Ontobee (https://ontobee.org/ontology/VBO), and BioPortal (https://bioportal.bioontology.org/ontologies/VBO). The v2024-11-06 version of VBO includes 19,541 terms representing breeds from vertebrate species, such as livestock (e.g., horse, cattle, chicken) and companion animals (e.g., dogs, cats), covering 49 species. VBO breed terms are classified under the general term 'Vertebrate breed' (VBO:0400000, Figure 1A) and grouped by specific animal species (e.g., 'Horse breed'), subspecies (e.g., 'Dog breed', Figure 1B) or genus (e.g., 'Cattle breed', Figure 2) based on community and expert usage and jargon. For example, 'Dog breed' (VBO:0400024) is "a breed of *Canis lupus familiaris*." This definition excludes wild species, such as *Canis rufus* (red wolf), *Canis latrans* (coyote), and *Canis lupus* (gray wolf). In contrast, 'Cattle breed' (VBO:0400020) is based on the name used in agriculture and defined at the genus level as "a vertebrate breed that is a *Bos*." 'Cattle breed' groups all breeds of *Bos* (NCBITaxon:9903) including *Bos indicus* (zebu cattle, NCBITaxon:9915), *Bos taurus* (cattle, NCBITaxon:9913), *Bos indicus* × *Bos taurus* (hybrid cattle, NCBITaxon:30522), and *Bos grunniens* (yak, NCBITaxon:30521), following animal science and veterinary community conventions.[11,15,24]

Each term in VBO is identified by a unique and permanent ID (Table 1). In addition, each term must have a unique label. However, breeds from different species often share the same name. For example, "Abyssinian" is the name for breeds of horse, cat, and donkey. In addition, some breeds are commonly called by names that can represent other types of entities: "Cyprus" is used to refer to breeds of cat, cattle, donkey, and goat but also to the country "Cyprus" (see section "DAD-IS is a major breed source in VBO"). To ensure that all term labels were unique, we concatenated the breed's most common name with their species' common name, following the format: `'Most common name (Species)'`, in which "Species" is the English language name (e.g., 'Cyprus (Cat)'). Distinguishable sub-breeds, strains, or varieties are also included in VBO and are related to the ontological parent breed using an *is_a* relation.[16,25,26] For example, 'Chihuahua, Long-Haired (Dog)' (VBO:0200339) and 'Chihuahua, Smooth-Haired (Dog)' (VBO:0200340) are subclasses of 'Chihuahua (Dog)' (VBO:0200338, Figure 1B).

Term metadata and provenance are provided for each VBO term. Metadata fields, definitions, and examples are provided in the VBO documentation.[14] Required term metadata include the most common name (a synonym tagged to indicate the name by which a breed is most commonly known), source (indicating provenance of the information), and contributor (Open Researcher and Contributor ID (ORCID)[27] of the curator(s) and/or expert(s) who contributed to the creation and/or revision of a VBO term) (Table 2). Additional metadata, such as other synonym(s), database cross-reference(s), breed codes, breed recognition status, breed domestication status, breed extinction status, and description of origin are included when available. Information from different sources might be discordant (e.g., breed recognition status by different registration bodies). We chose not to arbitrate and instead record all breed information, relying on provenance (e.g., source annotations) to guide users. For example, the 'Australian Mist (Cat)' (VBO:0100034) is a 'fully recognized breed' of the Governing Council of the Cat Fancy (GCCF), Rare and Exotic Feline Registry (REFR), The International Cat Association (TICA), and the World Cat Federation (WCF) and a 'not recognized breed' of the Fédération Internationale Féline (FIFe). Both recognition statuses are recorded in VBO (Table 2).

**VBO includes breed's foundation stock**

Breeds are sometimes created by crossing other breeds whose traits and/or pedigrees are desirable. For example 'Himalayan (Cat)' (VBO:0100117) was created from a cross of individuals from 'Siamese (Cat)' (VBO:0100221) and 'Persian (Cat)' (VBO:0100188) breeds (Figure 3). The animals that are the progenitors, or founders, of a breed are called "foundation stock";[28] they provide part of the underlying genetic base for a new distinct population. VBO provides information about breeds' foundation stock using the '*has foundation stock*' relation. This relation is defined as a relation between two distinct material entities (breeds or species), a descendant entity and an ancestor entity, in which the descendant entity is the result of mating, manipulation, or geographical or cultural isolation of the ancestor entity, therefore inheriting some of the ancestor's genetic material. Foundation stock can be one or more breeds, including wild animals represented by an NCBITaxon entity.

**DAD-IS is a major breed source in VBO**

Because FAO compiles and maintains a list of breeds reported by country-nominated National Coordinators from 182 countries, we partnered to include their DAD-IS breed list. The goal of the DAD-IS breed list is the management of animal genetic resources, focusing on the diversity of livestock breeds on the national, regional, and global levels, including the status of breeds regarding their risk of extinction. DAD-IS includes specific information related to its goals;[15] therefore, the corresponding breeds in VBO have unique associated metadata and semantic information representing this information, such as breed domestication status and breed extinction status.

Most records in DAD-IS represent breeds that have been reported by officially nominated FAO National Coordinators to exist in a specific country.[29] This concept, specific to DAD-IS, is represented in VBO using the '*breed reported in geographic location*' axiom indicating the geographical location where the breed was reported, using a Wikidata entry as the country identifier.[30] In addition, to ensure term label uniqueness, the naming conventions for these breeds follow the format `'Most common name, Country (Species)'`, in which "Country" and "Species" are the English language names (Figure 2). It is important to note that this concept of "breed that exists in a specific country as reported by FAO National Coordinators" is unique to DAD-IS and its goals. This concept is rarely used in contexts outside DAD-IS, the FAO, and related projects, and therefore the corresponding VBO terms should be used with caution.

**VBO is a community-driven resource**

Though maintained by the Monarch Initiative[13], VBO is a community resource that involves the participation of the community at large. Anyone can request changes or to add new breeds to the ontology and participate in discussions through the GitHub Issue Tracker (https://github.com/monarch-initiative/vertebrate-breed-ontology/issues).

# Discussion

VBO is a unique, open, community-driven ontology for vertebrate breeds, covering a broad scope of animals including livestock and companion animals, and encompassing breeds as defined by and in the context of international organizations and communities. VBO is a standard for breed terms, which supports data disambiguation and integration. Its hierarchical classification of concepts and defined relationships between concepts allow computational logical reasoning,[8,31] which can be leveraged in predictive tools. VBO is fundamental to the construction of veterinary clinical decision support tools that provide information about disease susceptibility in breeds and precision medicine tools to identify optimal treatments for individual animals. In addition, VBO can be leveraged for cross-species translational research[13,32] and conservation medicine work.[33] Here, we discuss the adoption of VBO for data annotation and disambiguation, ontology modeling decisions and their consequences for VBO usability, clinical and research applications, and upcoming improvements.

The same breed is often referenced using different names across (and sometimes within) veterinary databases and scientific reports, since no universal standard for breed names has yet been adopted in veterinary medicine. Using VBO IDs in these databases and reports disambiguates breed-related data.[34,35] For example, the information in Online Mendelian Inheritance in Animals (OMIA) has been rendered more interoperable by using VBO terms to specify breeds in which a trait, disorder, and/or a likely causal variant has been documented.[36]

We relied on external sources, such as international organizations, communities, and experts, to determine whether a term should be added to VBO. However, these external sources have specific purposes (e.g., breed competitions or breed diversity), and they often disagree on whether a group of animals should be recognized as a breed. For example, VeNom[12] lists 'Labradoodle (Dog)' (VBO:0200798) as a breed although it is not included in breed lists of canine registration bodies, such as the American Kennel Club, Fédération Cynologique Internationale, or United Kennel Club. We took an inclusive approach and created a VBO term when any sources described a breed, making VBO relevant to a broad range of use cases. We trust that the provenance of all pieces of information recorded in VBO will guide users in their decision whether or not to include particular VBO term(s) for their specific application.

Our ontology modeling decisions were driven by the need to address a broad range of use cases and warrant some discussion. The inclusion of species (and country, when applicable) in VBO term labels makes these labels unwieldy and cumbersome and somewhat represents the ontology hierarchy and axioms, which is not standard ontology practice. We investigated several approaches to create more user-friendly term labels. However, since the same common names can refer to breeds of different species, and no other specific breed attribute could be consistently identified, including species in the term label naming conventions was the most reliable solution to ensure term label uniqueness. We believe that the use of synonyms will help users identify and report their breeds of interest.

Based on our broad definition of vertebrate breed, breeds were created as subclasses of species, which are represented by the NCBITaxon entities; hence, we integrated VBO terms within the NCBITaxon hierarchy using the *is_a* relation. This has two main advantages: first, since NCBITaxon is an already accepted standard in ontologies and databases, it bridges VBO to other ontologies that use NCBITaxon in axioms (e.g., Mondo Disease Ontology that captures non-human animal diseases[37]). Second, we can leverage the species classification from NCBITaxon to autoclassify VBO terms through term axioms and logical definitions. For example, by defining 'Cattle breed' as "`'Vertebrate breed' and 'Bos'`", all VBO terms that are a subclass of the NCBITaxon are automatically classified as 'Cattle breed' (Figure 2). One should, however, be aware that the NCBITaxon hierarchy is motivated by evolutionary lineage[38] and, therefore, some classifications might be unintuitive to some users. For example, 'Chicken breed' (VBO:0400010) is defined as "a breed of Gallus gallus". In NCBITaxon, 'Gallus gallus' (NCBITaxon:9031) is a subclass of 'Phasianinae' (NCBITaxon:9072). Therefore, 'Chicken breed' is classified as a subclass of 'Pheasant breed' (VBO:0400037) since 'Pheasant breed' is defined as "a breed of Phasianinae" (Figure 4A). Similarly, 'Quail breed' (VBO:0001223) is a subclass of 'Partridge breed' (VBO:0400038) (Figure 4B). While we recognize that this classification is not intuitive to livestock experts, we trust that the use of synonyms and search engines will support the identification of VBO terms in these breed groups.

When NCBITaxonomy was translated into the Web Ontology Language (OWL) to be integrated into ontologies, NCBITaxon terms were defined to refer to classes of individual animals. VBO also refers to classes of breeds, which are subclasses of NCBITaxon classes. For example, "Snoopy is an instance of 'Beagle (Dog)' (VBO:0200131), which is a subclass of dog ('*Canis lupus familiaris*', NCBITaxon:9615)." While it is a crucial point for ontology developers to understand, this has minimal consequences for non-ontology creators. VBO terms can be used to refer to breeds in the same way that NCBITaxon terms are used to refer to genus, family, species, or subspecies.

Disease prediction tools are augmented by leveraging the computational logical reasoning of VBO, such as in the context of disease susceptibility in specific breeds. For example, it has been shown that Persian cats are susceptible to developing autosomal dominant polycystic kidney disease (ADPKD) (MONDO:1011054) due to a variant in the polycystin 1, transient receptor potential channel interacting gene (*PKD1*, NCBIGene:100144606).[39–41] The axioms in VBO indicate that 'Persian (Cat)' (VBO:0100188) is a foundation stock for 'Exotic Shorthair (Cat)' (VBO:0100096), which itself is a foundation stock for 'Foldex (Cat)' (VBO:0100099) (Figure 3). Based on these relationships, one could theorize that 'Exotic Shorthair (Cat)' and 'Foldex (Cat)' could also be susceptible to developing ADPKD; this has indeed been reported to be the case. It has also been shown that a variant in *PKD1* is associated with this susceptibility, as in 'Persian (Cat)'.[39–41] Similar predictions could guide disease predictions and treatment discovery that might be appropriate for some breeds but not others within the same species.

Breeds are often defined by their phenotypic features or traits, which are often associated with genetic components. For example, horse breeds with leopard complex spotting coat color pattern, such as 'Appaloosa (Horse)' (VBO:0000904), 'American Miniature Horse (Horse)' (VBO:0000896) and 'Knabstrupper (Horse)' (VBO:0001008), are susceptible to the disease congenital stationary night blindness (MONDO:1011255).[42] Therefore, any knowledge related to this disease (e.g., susceptibility) could be predicted to apply to all breeds sharing the leopard complex spotting coat color pattern. While not sufficient to make treatment decisions, this prediction informs veterinarians and researchers on avenues and directions for additional investigations.

VBO integrates extensive breed information and its provenance. While it has significant potential, additional enhancements would be beneficial. Many animal breeds have been historically based on conformation standards, including structure and appearance (e.g., coat color, hair length, size), as demonstrated by reports and text descriptions in international breed organizations and breed references.[18] Advancements in genomics bring new breed information and, in some cases, question the validity of the ancestry of these breeds.[34,43] In addition, the discovery of genetic components associated with trait aspects (e.g., variations determining cat tail length[44] and the effect of genotype on performance in racing 'Standardbred (Horse)' (VBO:0000899) and 'Scandinavian Coldblood Trotter (Horse)' (VBO:0017173)[45]) are changing how breeds are defined and how individuals are selected for breeding. Disparity between breeds defined by traits versus genetics can have a large impact on veterinary data and decisions. For example, treatment efficacy is affected by genetic factors.[3] Therefore, predicting treatment efficacy across breeds would be more accurate if breeds are related to each other based on genetics and not only based on traits. The introduction of relationships and classifications in VBO to specify how breeds are related (i.e., genetically versus phenotypically) will augment VBO's potential.

The majority of the VBO content was created by manually reviewing existing lists of breeds and classifying them according to their species. However, the number of breeds and the wealth of available information is massive. In addition, the majority of available information is in free text format in databases, publications, health records, and books.[18,46] Therefore, the manual review used to create VBO is neither sustainable nor scalable. We will leverage the advances of large language models in ontology curation tools[47] to continue to populate VBO with relevant information. In addition, communities often classify breeds based on their usage, such as "used for meat production", "used for milk production", "draft horse", "sighthound", etc. Such classification would also be relevant to add to future VBO versions.

      VBO is a powerful tool to improve veterinary clinical and research efforts moving forward. Veterinary EHR is the ideal environment to implement VBO as it directly interacts with clinical decision-support tools. An achievable first step would be annotating research artifacts, including journal articles and datasets, with VBO IDs. The VBO IDs would aid in collating studies performed in animals of the same breed for data integration in systematic reviews and meta-analyses and help overcome the limitations of small sample sizes in most prospective veterinary studies.[48] VBO is built by the community for the community and is a blueprint and a significant first step toward achieving data harmonization in veterinary medicine.

**Table 1: Minimum Information for the Reporting of an Ontology (MIRO) for VBO.**

| **A. The basics** | |
|---|---|
| A.1 Ontology name | Vertebrate Breed Ontology (VBO)<br>Version in this manuscript: v2024-11-06 |
| A.2 Ontology owner | Monarch Initiative (https://monarchinitiative.org) |
| A.3 Ontology license | CC-BY 4.0 |
| A.4 Ontology URL | http://purl.obolibrary.org/obo/vbo.owl |
| A.5 Ontology repository | https://github.com/monarch-initiative/vertebrate-breed-ontology |
| A.6 Methodological framework | Breed lists were collected from international breed organizations and communities. These lists were manually curated and integrated such that the same breed concept and its information were represented by a single VBO term. Term classification based on species (NCBITaxon) was also done manually. |
| **B. Motivation** | |
| B.1 Need | Single source for vertebrate breed name and related information, representing a broad range of species and breeds, available open access, and including provenance for the information. |
| B.2 Competition | Livestock Breed Ontology (LBO) is a resource for livestock breeds. LBO is, however, limited to livestock breeds and does not include companion animals (such as cat and dog breeds). In addition, many new livestock breeds (e.g., some breeds reported in DAD-IS) are also out of scope in VBO. |
| B.3 Target audience | - Databases, such as OMIA.<br>- Any sources containing breed information; for example veterinary EHR.<br>- Publications, in order to disambiguate breeds and enable data curation and integration. |
| **C. Scope, requirements, development community (SRD)** | |
| C.1 Scope and coverage | All vertebrate breeds, including sub-breeds, varieties, strains, etc. |
| C.2 Development community | The content of the ontology was initially created from international breed organizations and community lists.<br>Additional VBO terms and breed information are driven by user requests and breed sources update. |
| C.3 Communication | - GitHub issue tracker: https://github.com/monarch-initiative/vertebrate-breed-ontology/issues<br>- Mailing list: vbo-users@tislab.org (subscribe by sending an email to: vbo-users+subscribe@tislab.org) |
| **D. Knowledge acquisition (KA)** | |
| D.1 Knowledge acquisition method | Breed lists were collected from international breed organizations and communities and manually curated, with consultation with animal |

| | experts. Review and verification happen with targeted reviews involving experts and via user requests. |
|---|---|
| D.2 Source knowledge location | Sources where the breed knowledge was gathered can be found here: https://monarch-initiative.github.io/vertebrate-breed-ontology/general/general/ |
| D.3 Content selection | User requests for new breeds and updates to the existing ontology are given priority.<br><br>While synchronization of information with the original sources, as well as the addition of new breed sources, are also of high priority, these are addressed on a per-available-ontology-editor-resources basis. |
| **E. Ontology content** | |
| E.1 Knowledge representation language | The Web Ontology Language (OWL) is used, as it is more expressive and allows axioms such as "*of breed reported in* value [Wikidata ID for country]", which are necessary for the majority of the breeds from the DAD-IS list.<br><br>.obo and .json versions of the ontology are also available; however, these formats do not include all information included in the .owl format (e.g., "*breed reported in geographic location*" axioms). |
| E.2 Development environment | The Ontology Development Kit (ODK, https://github.com/INCATools/ontology-development-kit) was used to create and is used to maintain VBO. |
| E.3 Ontology metrics | As of version (v2024-11-06):<br>- 19,541 Classes<br>- 118 Object properties<br>- 261,945 Axioms (55,012 Logical axioms) |
| E.4 Incorporation of other ontologies | - NCBITaxon (though not an ontology *per se*) is at the basis of VBO.<br>- Relation Ontology (RO) is used for relationships.<br>- Core Ontology for Biology and Biomedicine (COB) is used as an upper-level ontology.<br>These are updated before each release, which aims to be monthly. |
| E.5 Entity naming convention | Naming conventions for term labels follow the format:<br>- `'Most common name (Species)'`<br>  in which Species is the English name (e.g. 'Chihuahua (Dog)' (VBO:0200338)).<br>- `'Most common name, Country (Species)'`<br>  in which country and species are the English names. (e.g., 'Jersey Giant, Canada (Chicken)' (VBO:0006068)). This format is used for breeds reported to exist in specific geographic locations by country-nominated FAO National Coordinators in DAD-IS.<br>More information about naming conventions here: https://monarch-initiative.github.io/vertebrate-breed-ontology/ontologymodeling/term-labels-naming-conventions/ |
| E.6 Identifier generation policy | http://purl.obolibrary.org/obo/VBO_ followed by 7 digits, e.g., http://purl.obolibrary.org/obo/VBO_0200338 |
| E.7 Entity metadata policy | The following metadata is required:<br>- ID<br>- Term label<br>- Most common name<br>- Source<br>- Contributor<br>Other metadata is optional, for example:<br>- Synonyms<br>- Database cross-reference<br>- Breed code<br>- …<br>A full list of metadata and their descriptions can be found in our documentation:<br>https://monarch-initiative.github.io/vertebrate-breed-ontology/ontologymodeling/metadata/ |
| E.8 Upper ontology | Core Ontology for Biology and Biomedicine (COB) |
| E.9 Ontology relationships | RO relations are used. The main relation used is **is_a**.<br><br>Additional relations specific to breed information were created in VBO. For example:<br>- **has foundation stock** (VBO:0300019)<br>- **breed reported in geographic location** (VBO:0300020)<br><br>See full explanation of relationships:<br>https://monarch-initiative.github.io/vertebrate-breed-ontology/ontologymodeling/axioms/<br><br>We plan on submitting all VBO-created relations to RO. If in scope, and new relations are created in RO, these will replace the current VBO relations. |
| E.10 Axiom pattern | See documentation:<br>https://monarch-initiative.github.io/vertebrate-breed-ontology/general/general/ |
| E.11 Dereferenceable IRI | Standard prefix: VBO |
| **F. Managing change** | |
| F.1 Sustainability plan | VBO is actively developed and maintained by the members of the Monarch Initiative and the OMIA team. Training sessions and new ontology editing tools are available at the OBO Academy[49] and in the works to empower more users (who are not necessarily ontologists) to edit the ontology directly. |
| F.2 Entity deprecation strategy | Term deprecation happens when:<br>- A VBO term represents a concept that never existed (i.e., created by |
| | mistake). In this case, the term is obsoleted.<br>  - Their IDs are maintained with the annotation 'owl:deprecated': true.<br>  - The obsoletion reason is recorded using the OBO Metadata Ontology (OMO) term 'domain entity does not exist' (OMO:0001002).<br>  - A link to the issue tracker discussing this obsoletion is recorded using 'term tracker item' (IAO:0000233) annotation.<br>- A VBO term represents the same breed concept as another VBO term (i.e., concepts are duplicated). In this case, the terms are merged.<br>  - The ID of the merged term, i.e., that is obsoleted, is maintained with the annotation 'owl:deprecated': true.<br>  - The obsoletion reason is recorded using the OMO term "term merged".<br>  - The annotation "replaced by" indicates the VBO ID of the term into which it was merged.<br>  - A link to the issue tracker discussing this merge is recorded using 'term tracker item' (IAO:0000233) annotation. |
| F.3 Versioning policy | We aim to release a new ontology version every other month. |
| **G. Quality Assurance (QA)/Quality Control (QC)** | |
| G.1 Testing | We use the general QA/QC included in the ODK framework. In addition, we are working on creating new QC tests specific to VBO. |
| G.2 Evaluation | Currently, the ontology is evaluated by:<br>- Users who use VBO in data annotations and check that all required VBO terms exist.<br>- Ontology developers who review the ontology by comparing it to major breed sources (e.g., Mason's World Dictionary of Livestock Breeds, Types and Varieties,[46] and Mason's World Encyclopedia of Livestock Breeds and Breeding[18]).<br><br>Evaluation using data integration and competency questions will be possible once VBO has been fully adopted by multiple databases. |
| G.3 Examples of use | - https://omia.org/OMIA000515/9685/<br>- https://omia.org/OMIA001081/9615/<br>- https://www.cellosaurus.org/CVCL_2L83<br>- https://www.cellosaurus.org/CVCL_L309 |
| G.4 Institutional endorsement | VBO has been accepted to the OBO foundry.[21] |
| G.5 Evidence of use | - OMIA (https://omia.org/home/) records breed information and uses VBO ID to refer to breeds.<br>- Cellosaurus (https://www.cellosaurus.org/) uses VBO ID to refer to breeds.<br>- Selected publications using VBO ID to refer to breeds.[34,35] |

*Note:* This table was created based on the MIRO guidelines as described in Matentzoglu et al.[22]

Abbreviations: NCBITaxon, National Center for Biotechnology Information Taxonomy; DAD-IS, Domestic Animal Diversity Information System; OMIA, Online Mendelian Inheritance in Animals; EHR, Electronic Health Record; ID = identification; OBO, Open Biomedical Ontologies. All other abbreviations are defined in the table.

**Table 2: Selected metadata for 'Australian Mist (Cat)' (VBO:0100034).**

| Metadata | Annotations | Sources |
|---|---|---|
| has exact synonym most common name | Australian Mist | https://fifeweb.org/cats/ems-system/ https://en.wikipedia.org/wiki/List_of_cat_breeds#Breeds https://wcf.de/en/wcf-ems-code/ https://www.gccfcats.org/getting-a-cat/choosing/cat-breeds/ https://www.rareexoticfelineregistry.com/breed-recognition/ https://www.tica.org/breeds/browse-all-breeds https://www.worldcatcongress.org/wp/cat_breed_comp_aum.php |
| breed code | AMS | https://www.rareexoticfelineregistry.com/breed-recognition/ |
|  | AUM | https://www.tica.org/breeds/browse-all-breeds https://wcf.de/en/wcf-ems-code/ https://www.gccfcats.org/getting-a-cat/choosing/cat-breeds/ |
|  | AUM non | https://fifeweb.org/cats/ems-system/ |
| breed recognition status | fully recognized breed | https://wcf.de/en/wcf-ems-code/ https://www.gccfcats.org/wp-content/uploads/2024/04/SOP.final_.April2024.pdf https://www.rareexoticfelineregistry.com/breed-recognition/ https://www.tica.org/breeds/browse-all-breeds |
|  | not recognized breed | https://fifeweb.org/cats/ems-system/ |
| contributors | https://orcid.org/0000-0002-1628-7726 https://orcid.org/0000-0002-4142-7153 https://orcid.org/0000-0002-5002-8648 https://orcid.org/0000-0002-5520-6597 https://orcid.org/0000-0002-9178-3965 |  |

A  Vertebrata <vertebrates>
- Vertebrate breed
  - Bird breed
  - Bovine breed
    - American Bison breed
    - Buffalo breed
    - Cattle breed
  - Camel breed
  - Cat breed
  - Deer breed
  - Dog breed
  - Equid breed
    - Ass breed
    - Horse breed
  - Fish breed
  - Frog breed
  - Goat breed
  - Golden hamster breed
  - Guinea pig breed
  - Pig breed
  - Rabbit breed
  - Sheep breed
  - South American Camelid breed
    - Alpaca breed
    - Guanaco breed
    - Llama breed
    - Vicuña breed

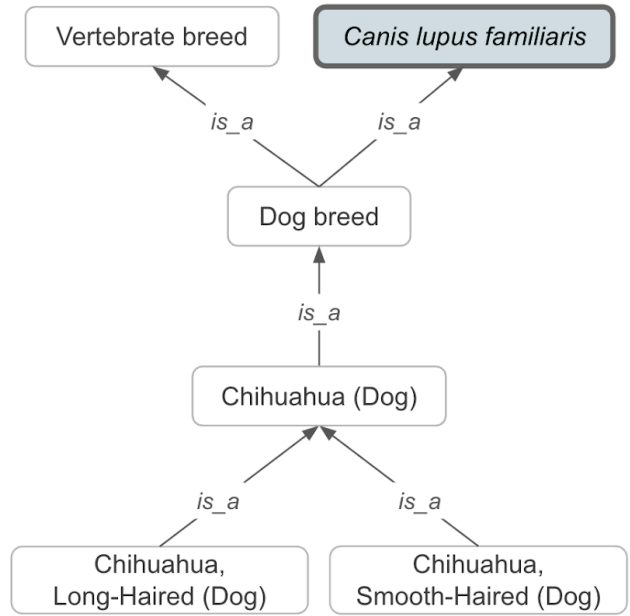

**Figure 1: Classification of vertebrate breeds. (A)** High-level classification based on species (e.g., 'Dog breed') and community usage (e.g., 'Cattle breed"). **(B)** VBO representation of Chihuahua dog breeds in VBO. 'Chihuahua (Dog)' is a subclass of 'Dog breed', itself a subclass of 'Vertebrate Breed' and '*Canis lupus familiaris'.* 'Chihuahua, Long-Haired (Dog)' and 'Chihuahua, Smooth-Haired (Dog)' are subclasses of 'Chihuahua (Dog)', since they are more specific instances of 'Chihuahua (Dog)'.

Terms from the NCBITaxon hierarchy are shown in thick-border boxes. Arrows represent *is_a* relation. Some relations and VBO terms are not displayed in this figure for clarity. All VBO and NCBITaxon IDs are reported in the Supplemental File.

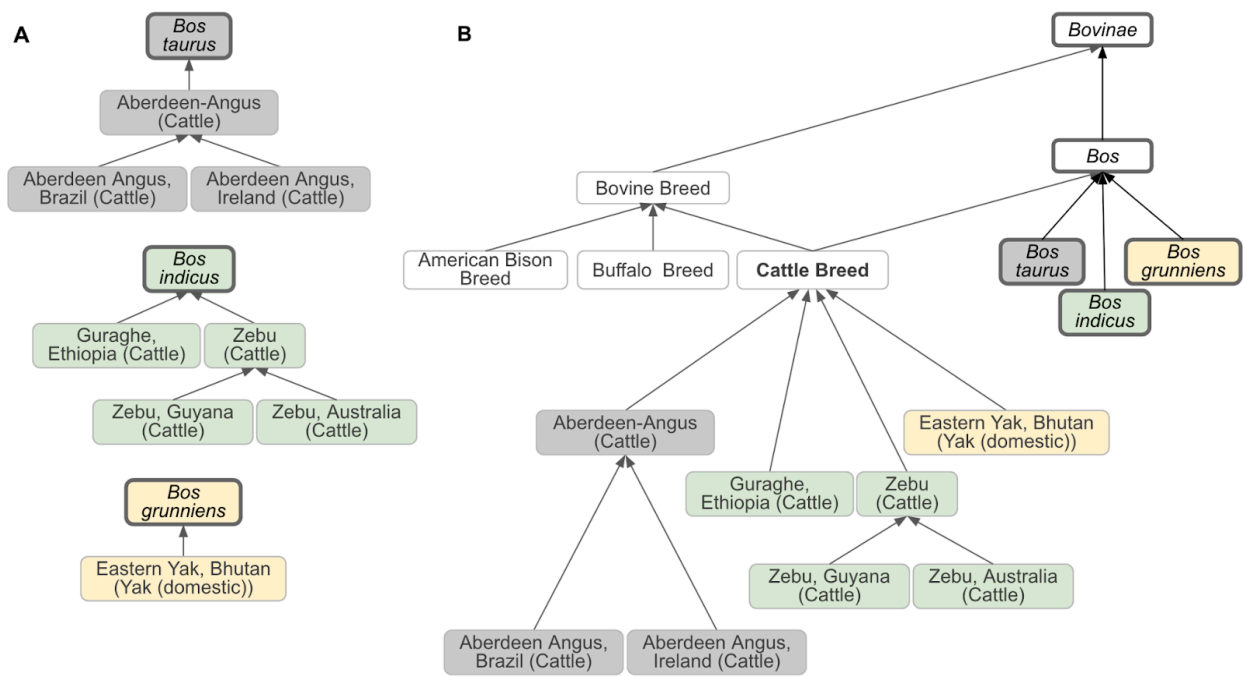

**Figure 2: Classification of bovine breeds. (A)** Relation between selected cattle breeds and their NCBITaxon species of *Bos taurus, Bos indicus, and Bos grunniens*. Each VBO term is related to an NCBITaxon. Breeds defined as having been reported in a specific country by National Coordinators in DAD-IS are either direct subclasses of their corresponding NCBITaxon or subclasses of other breeds (and therefore inherit the NCBITaxon subclass). Direct subclasses of NCBITaxon shown in this figure are 'Guraghe, Ethiopia (Cattle)' and 'Eastern Yak, Bhutan (Yak (domestic))'. Subclasses of other breeds shown in this figure are as follows: 'Aberdeen Angus, Brazil (Cattle)' and 'Aberdeen Angus, Ireland (Cattle)', subclasses of 'Aberdeen-Angus (Cattle)'; and 'Zebu, Guyana (Cattle)' and 'Zebu, Australia (Cattle)', subclasses of 'Zebu (Cattle)'. **(B)** 'Cattle Breed' is defined as "A breed of *Bos*" and is, therefore, a subclass of *Bos*. *Bos* encompasses *Bos taurus, Bos indicus, and Bos grunniens.* As a consequence, all breeds of these NCBITaxon species (as shown in A) are classified under 'Cattle Breed'. Similarly, 'Bovine Breed' being defined as "A breed of *Bovinae"* encompasses 'Cattle Breed' (of taxon *Bos)*, 'American Bison Breed' (of taxon *Bison bison)*, and 'Buffalo Breed' (of taxon *Bubalus bubalis),* since *Bos*, *Bison bison*, and *Bubalus bubalis* are subclasses of *Bovinae* (not shown). The relation between the breed(s) and their corresponding NCBITaxon is not shown for clarity.

Terms from the NCBITaxon hierarchy are shown in thick-border boxes. Boxes of the same color represent a unique NCBITaxon and its ontological children. Arrows represent *is_a* relation. Some relations and VBO terms are not displayed in this figure for clarity. All VBO and NCBITaxon IDs are reported in the Supplemental File.

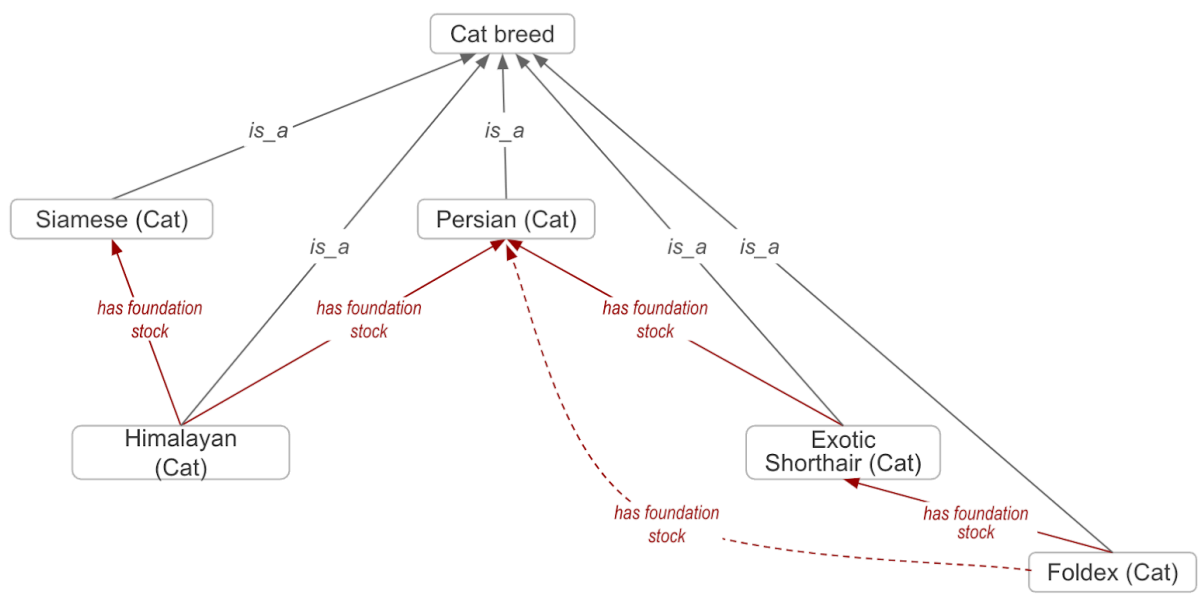

**Figure 3: Breeds are related to their progenitors via the *has foundation stock* relation.**
'Himalayan (Cat)' has progenitors *'Siamese (Cat)'* and 'Persian (Cat)'; therefore this breed is related to them via the *has foundation stock* relation. Due to the transitive property of the *has foundation stock* relation, terms inherit the *has foundation stock* from their progenitor(s). For example, 'Foldex (Cat)' *has foundation stock* 'Exotic Shorthair (Cat)', which itself *has foundation stock* 'Persian (Cat)'. Therefore, it can be inferred that 'Foldex (Cat)' *has foundation stock* 'Persian (Cat)' (dashed arrow).
Full arrows represent *is_a* relation; dashed arrows represent *has foundation stock* relation. Some relations and VBO terms are not displayed in this figure for clarity. All VBO and NCBITaxon IDs are reported in the Supplemental File.

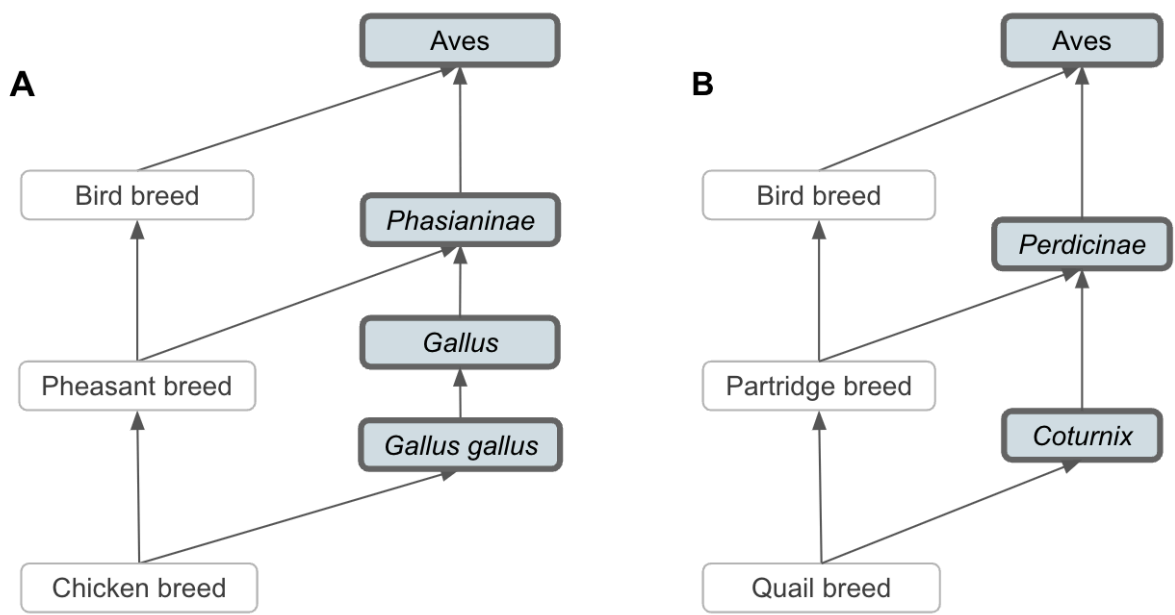

**Figure 4: NCBITaxon is leveraged to autoclassify VBO terms.**
**(A)** NCBITaxon classifies '*Gallus gallus*' as a subclass of *'Phasianinae'*. Since 'Chicken breed' is a subclass of '*Gallus gallus*', and 'Pheasant breed' is a subclass of *'Phasianinae'*, 'Chicken breed' is autoclassified as a subclass of 'Pheasant breed'. **(B)** NCBITaxon classifies '*Coturnix*' as a subclass of '*Perdicinae*'. Since 'Quail breed' is a subclass of '*Coturnix*', and 'Partridge breed' is a subclass of *'Perdicinae'*, 'Quail breed' is autoclassified as a subclass of 'Partridge breed'.
Terms from the NCBITaxon hierarchy are shown in thick-border boxes. Arrows represent *is_a* relation. Some relations, NCBITaxon, and VBO terms are not displayed in this figure for clarity. All VBO and NCBITaxon IDs are reported in the Supplemental File.

**Supplementary Table: Identifiers reported in this publication.**

| Categories | Labels | IDs | Location in the document |
|---|---|---|---|
| VBO term-breed | Aberdeen Angus, Brazil (Cattle) | VBO:0002150 | Fig.2 |
| VBO term-breed | Aberdeen Angus, Ireland (Cattle) | VBO:0002169 | Fig.2 |
| VBO term-breed | Aberdeen-Angus (Cattle) | VBO:0000090 | Fig.2 |
| VBO term-breed | American Miniature Horse (Horse) | VBO:0000896 | Text |
| VBO term-breed | Appaloosa (Horse) | VBO:0000904 | Text |
| VBO term-breed | Australian Mist (Cat) | VBO:0100034 | Text, Fig.4 |
| VBO term-breed | Beagle (Dog) | VBO:0200131 | Text |
| VBO term-breed | Chihuahua (Dog) | VBO:0200338 | Fig.1 |
| VBO term-breed | Chihuahua (Dog) | VBO:0200338 | Table 1 |
| VBO term-breed | Chihuahua, Long-Haired (Dog) | VBO:0200339 | Text, Fig.1 |
| VBO term-breed | Chihuahua, Smooth-Haired (Dog) | VBO:0200340 | Text, Fig.1 |
| VBO term-breed | Cyprus (Cat) | VBO:0100081 | Text |
| VBO term-breed | Eastern Yak, Bhutan (Yak (domestic)) | VBO:0016815 | Fig.2 |
| VBO term-breed | Exotic Shorthair (Cat) | VBO:0100096 | Fig.3 |
| VBO term-breed | Foldex (Cat) | VBO:0100099 | Fig.3 |
| VBO term-breed | Guraghe, Ethiopia (Cattle) | VBO:0004734 | Fig.2 |
| VBO term-breed | Himalayan (Cat) | VBO:0100117 | Text, Fig.3 |
| VBO term-breed | Jersey Giant, Canada; Chicken | VBO:0006068 | Table 1 |
| VBO term-breed | Knabstrupper (Horse) | VBO:0001008 | Fig.5 |
| VBO term-breed | Labradoodle (Dog) | VBO:0200798 | Text |
| VBO term-breed | Lakenvelder, Belgium (Cattle) | VBO:0002866 | Fig.5 |
| VBO term-breed | Persian (Cat) | VBO:0100188 | Text, Fig.3 |
| VBO term-breed | Plott Hound (Dog) | VBO:0201023 | Text |
| VBO term-breed | Scandinavian Coldblood Trotter (Horse) | VBO:0017173 | Text |
| VBO term-breed | Siamese (Cat) | VBO:0100221 | Text, Fig.3 |
| VBO term-breed | Standardbred (Horse) | VBO:0000899 | Text |
| VBO term-breed | Zebu (Cattle) | VBO:0017417 | Fig.2 |
| VBO term-breed | Zebu, Australia (Cattle) | VBO:0004402 | Fig.2 |
| VBO term-breed | Zebu, Guyana (Cattle) | VBO:0004839 | Fig.2 |
| VBO term-classification | Alpaca breed | VBO:0000038 | Fig.1 |
| VBO term-classification | American bison breed | VBO:0000041 | Fig.1, Fig.2 |
| VBO term-classification | Ass breed | VBO:0400005 | Fig.1 |
| VBO term-classification | Bird breed | VBO:0400006 | Text, Fig.1 |
| VBO term-classification | Bovine breed | VBO:0400019 | Fig.1, Fig.2 |
| VBO term-classification | Buffalo breed | VBO:0000068 | Fig.1, Fig.2 |
| VBO term-classification | Camel breed | VBO:0400022 | Fig.1 |
| VBO term-classification | Cat breed | VBO:0400018 | Fig.1, Fig.3 |
| VBO term-classification | Cattle breed | VBO:0400020 | Text, Fig.1, Fig.2 |
| VBO term-classification | Chicken breed | VBO:0400010 | Text, Fig. 5 |
| VBO term-classification | Deer breed | VBO:0400023 | Fig.1 |
| VBO term-classification | Dog breed | VBO:0400024 | Text, Fig.1 |
| VBO term-classification | Equid breed | VBO:0400033 | Fig.1 |
| VBO term-classification | Goat breed | VBO:0400025 | Fig.1 |
| VBO term-classification | Guanaco breed | VBO:0000882 | Fig.1 |
| VBO term-classification | Guinea pig breed | VBO:0400026 | Fig.1 |

| VBO term-classification | Horse breed | VBO:0000931 | Fig.1 |
| --- | --- | --- | --- |
| VBO term-classification | Llama breed | VBO:0001098 | Fig.1 |
| VBO term-classification | Partridge breed | VBO:0400038 | Text, Fig. 5 |
| VBO term-classification | Pheasant breed | VBO:0400037 | Text, Fig. 5 |
| VBO term-classification | Pig breed | VBO:0001199 | Fig.1 |
| VBO term-classification | Quail breed | VBO:0001223 | Text, Fig. 5 |
| VBO term-classification | Rabbit breed | VBO:0400029 | Fig.1 |
| VBO term-classification | Sheep breed | VBO:0400030 | Fig.1 |
| VBO term-classification | South American camelid breed | VBO:0400032 | Fig.1 |
| VBO term-classification | Vertebrate Breed | VBO:0400000 | Text, Fig.1 |
| VBO term-classification | Vicuña breed | VBO:0001721 | Fig.1 |
| Taxon | Aves | NCBITaxon:8782 | Text |
| Taxon | *Bos* | NCBITaxon:9903 | Text, Fig.2 |
| Taxon | *Bos grunniens* | NCBITaxon:30521 | Fig.2 |
| Taxon | *Bos indicus* | NCBITaxon:9915 | Fig.2 |
| Taxon | *Bos indicus × Bos taurus* | NCBITaxon:30522 | Text |
| Taxon | *Bos taurus* | NCBITaxon:9913 | Fig.2 |
| Taxon | *Bovinae* | NCBITaxon:27592 | Fig.2 |
| Taxon | *Canis lupus familiaris* | NCBITaxon:9615 | Fig.1 |
| Taxon | *Coturnix* | NCBITaxon:9090 | Fig.5 |
| Taxon | *Gallus* | NCBITaxon:9030 | Fig.5 |
| Taxon | *Gallus gallus* | NCBITaxon:9031 | Text, Fig. 5 |
| Taxon | *Perdicinae* | NCBITaxon:466544 | Fig.5 |
| Taxon | *Phasianinae* | NCBITaxon:9072 | Text, Fig. 5 |

| Taxon | Vertebrata <vertebrates> | NCBITaxon:7742 | Fig.1 |
|---|---|---|---|
| Gene | PKD1, polycystin 1, transient receptor potential channel interacting, Felis catus (domestic cat) | NCBIGene:100144606 | Text |
| Disease | congenital stationary night blindness, TRPM1-related, horse | MONDO:1011255 | Text |
| Disease | polycystic kidney disease, domestic cat | MONDO:1011054 | Text |
| Breed status | domestication status | VBO:0300005 | Text |
| Breed status | extinction status | VBO:0300009 | Text |
| Breed status | fully recognized breed | VBO:0300002 | Text |
| Breed status | not recognized breed | VBO:0300004 | Text |
| relation | *breed reported in geographic location* | VBO:0300020 | Text, Table 1 |
| relation | *has foundation stock* | VBO:0300019 | Text, Table 1 |

Note that the category "VBO term-breed" includes sub-breed and variety.